# Nonlocal Dispersion Cancellation for Three or More Photons

I. C. Nodurft[1], S. U. Shringarpure[1], B. T. Kirby[2], T.B. Pittman[1], and J. D. Franson[1]
[1]Physics Department, University of Maryland Baltimore County, Baltimore, Maryland 21250 USA
[2]Army Research Laboratory, Adelphi, Maryland 20783 USA

The entanglement of quantum systems can produce a variety of nonclassical effects that have practical applications in quantum information science. One example of this is nonlocal dispersion cancellation, in which the effects of dispersion on one photon can be canceled out by the dispersion experienced by a second photon at a distant location. In this paper, we extend the analysis of nonlocal dispersion cancellation to three or more photons. We find that energy-time entanglement of three or more photons can lead to a complete or partial cancellation of dispersion depending on the experimental conditions. These results may be useful in implementing quantum key distribution in networks with three or more nodes.

## I. Introduction

A short classical pulse of light propagating through a dispersive medium will become broadened, which can introduce a significant uncertainty in the time at which it will be detected. Two classical pulses traveling in two different media will be broadened independently, with a resulting increase in the uncertainty in their relative detection times. In contrast, two photons that are entangled in energy and time [1-2] can propagate through two different media in such a way that the dispersion experienced in one medium is cancelled out by the dispersion in the other medium [3-23]. In this paper, we extend the theory of nonlocal dispersion cancellation to three or more photons and show that complete or partial cancellation of dispersion can occur, depending on the experimental arrangement.

Nonlocal dispersion cancellation has a number of potential applications in quantum key distribution (QKD) or quantum networks, where the data rate can be limited by the effective pulse width. The reduced timing uncertainties are especially important for QKD systems based on nonlocal interferometry [1,2], where the difference in interferometer path lengths must be larger than the effective width of the wave packets. In addition, nonlocal dispersion cancellation itself can be used as the basis for quantum key distribution [24-27]. Roughly speaking, the presence of an eavesdropper will destroy the dispersion cancellation, which can be detected by the system. Nonlocal dispersion cancellation can also be employed for clock synchronization in a protocol that is resistant to pulse distortions caused in transit [28-30]. Biomedical imaging applications have also made use of nonlocal dispersion cancellation to improve the quality of the images [31-34]. We expect that the extension of nonlocal dispersion cancellation to higher numbers of photons will also have potential applications, especially for quantum networks with three or more nodes.

We will consider the tripartite entangled state created by a $\chi^{(3)}$ nonlinear crystal to study nonlocal dispersion cancellation for the three-photon case. A similar approach allows us to extend the results to larger photon numbers. Similar results can also be obtained using two cascaded $\chi^{(2)}$ crystals [35]. Three-photon entanglement has previously been used in other applications, such as nonlocal interferometry [35-37].

The paper is organized as follows. Section II calculates the effects of nonlocal dispersion cancellation for the three-photon entangled state created from a single $\chi^{(3)}$ down conversion process. Section III calculates the corresponding dispersion for three classical pulses of light. The classical and quantum-mechanical results are compared in Section IV. Section V extends the previous results for the three-photon case to higher photon numbers. Section VI provides a summary and conclusion. Additional details are provided in the Appendix.

## II. Three-photon dispersion cancellation

The most straightforward method for creating tripartite energy-time entangled photon states is through parametric down conversion [38,39]. In this section, we consider the generation of three entangled photons using a single down-conversion process in a $\chi^{(3)}$ nonlinear crystal as illustrated in Fig. 1. This process converts a high energy pump photon into three secondary photons with lower energies. From energy conservation, the sum of the frequencies of the three secondary photons must equal that of the pump photon, but in general their frequencies need not be equal. As is the case for two photons, the resulting three-photon state is entangled both in energy and time.

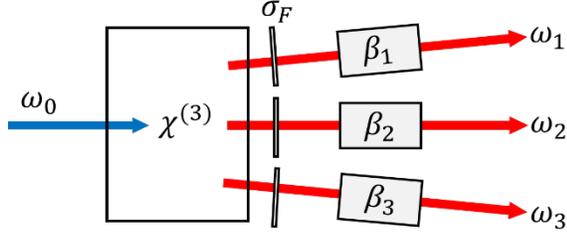

**Figure 1:** An entangled three-photon source using a $\chi^{(3)}$ nonlinear crystal pumped by a laser at a frequency of $\omega_0$. Each photon passes through a filter with bandwidth $\sigma_F$ before propagating through separate media with dispersion coefficients given by $\beta_1$, $\beta_2$, and $\beta_3$.

In the absence of any dispersion and in the limit of large bandwidths, the three photons would be detected at the same time if they travel equal distances to the detectors. That will no longer be the case in general in the presence of dispersion, and we calculate the probability distribution for the three photon detection times. The dispersion coefficients in the three media where the photons propagate will be denoted $\beta_1$, $\beta_2$, and $\beta_3$, while the propagation distances will be denoted $x_1$, $x_2$, and $x_3$.

The most general form of the initial state of the three down-converted photons is given by [35,38]

$$|\psi\rangle = \int d\omega_1 d\omega_2 d\omega_3\, g(\omega_1,\omega_2,\omega_3) \hat{a}^\dagger_{k_1} \hat{b}^\dagger_{k_2} \hat{c}^\dagger_{k_3} |0\rangle. \quad (1)$$

Here $\omega_1$, $\omega_2$, and $\omega_3$ are the angular frequencies of the photons while the corresponding photon creation operators will be denoted by $\hat{a}^\dagger_{k_1}$, $\hat{b}^\dagger_{k_2}$, and $\hat{c}^\dagger_{k_3}$. The function $g(\omega_1,\omega_2,\omega_3)$ is determined by the nonlinear crystal and the phase matching conditions. We will assume that the three photons pass through Gaussian filters whose bandwidths are sufficiently narrow that the function $g(\omega_1,\omega_2,\omega_3)$ can be approximated by the product of three Gaussians and a Dirac delta function for energy conservation. The state of the system after the three filters can then be written in the form

$$|\psi\rangle = c \int_0^{\omega_0} d\omega_1 \int_0^{\omega_0 - \omega_1} d\omega_2\, f_1 f_2 f_3\, \hat{a}^\dagger_{k_1} \hat{b}^\dagger_{k_2} \hat{c}^\dagger_{k_3} |0\rangle, \quad (2)$$

where $c$ is a constant and

$$f_i = \frac{1}{\sqrt{2\pi}\sigma_F} e^{-(\omega_i - \omega_F)^2/2\sigma_F^2}. \quad (3)$$

Here $\sigma_F$ represents the bandwidth of the filters, which are all assumed to be the same with central frequencies $\omega_F = \omega_0/3$. Eq. (2) makes use of the fact that the sum of the three frequencies must equal $\omega_0$.

It will be convenient to introduce three new variables $\epsilon_1$, $\epsilon_2$, and $\epsilon_3$ defined in such a way that

$$\omega_1 = \frac{\omega_0}{3} + \epsilon_1,$$
$$\omega_2 = \frac{\omega_0}{3} + \epsilon_2, \quad (4)$$
$$\omega_3 = \frac{\omega_0}{3} + \epsilon_3,$$

where $\epsilon_3 = -\epsilon_1 - \epsilon_2$. This leaves two independent variables $\epsilon_1$ and $\epsilon_2$.

In the Heisenberg picture, the positive frequency component of the electric field operator for each photon is given by

$$\hat{E}^+(x_i, t_i) = i \sum_{\omega_i} \left(\frac{2\pi\hbar\omega_i}{V}\right)^{1/2} e^{i(k_i x_i - \omega_i t_i)} \hat{a}_{k_i}. \quad (5)$$

The constant $V$ is the volume corresponding to the use of periodic boundary conditions while the wave numbers $k_i$ are a function of $\omega_i$ in a dispersive medium. The negative frequency component of the field operator is the Hermitian conjugate of the positive frequency component. We can define an effective wave function $\psi$ given by

$$\psi(x_1,x_2,x_3,t_1,t_2,t_3) \equiv \langle 0| \hat{E}^+(x_1,t_1)\hat{E}^+(x_2,t_2)\hat{E}^+(x_3,t_3)|\psi\rangle. \quad (6)$$

Inserting the change of variables in Eq. (4) into Eq. (2) and converting the sums to integrals gives

$$\psi = c' \int_{-\omega_0/3}^{2\omega_0/3} d\epsilon_1 \int_{-\omega_0/3}^{2\omega_0/3 - \epsilon_1} d\epsilon_2\, f_1 f_2 f_3 \\ \times e^{i\left[(k_1 x_1 + k_2 x_2 + k_3 x_3) - (\omega_1 t_1 + \omega_2 t_2 + \omega_3 t_3)\right]}. \quad (7)$$

Here $c'$ is a constant and the factors of $\omega_i$ in the electric field operators have been approximated by the central filter frequency $\omega_F$, which is valid when the width of the filters is sufficiently narrow.

As usual, the wavenumbers $k_i$ can be expanded in a Taylor series around the central frequency $\omega_F$:

$$k_i(\omega_i) = k_{F_i} + \alpha_i(\omega_i - \omega_F) + \beta_i(\omega_i - \omega_F)^2 \\ = k_{F_i} + \alpha_i \epsilon_i + \beta_i \epsilon_i^2. \quad (8)$$

Here we have assumed that the filters are sufficiently narrow that third and higher order terms can be neglected. The coefficients $\alpha_i$ of the first order terms are related to the group velocities whereas the coefficients $\beta_i$ of the second order terms give rise to dispersion.

Substituting Eq. (8) into Eq. (7) and extending the integrals to infinity under the assumption that the filter bandwidths are narrow compared to $\omega_0/3$ gives

$$\psi(t_1,t_2,t_3) = c' e^{i\left[k_{F_1}x_1 + k_{F_2}x_2 + k_{F_3}x_3 - (\alpha_1 x_1 + \alpha_2 x_2 + \alpha_3 x_3)\omega_0/3\right]} \\ \times \int_{-\infty}^{\infty} d\epsilon_1 \int_{-\infty}^{\infty} d\epsilon_2 \; f_1 f_2 f_3 \; e^{-i\left[\omega_1(t_1-\alpha_1 x_1)+\omega_2(t_2-\alpha_2 x_2)+\omega_3(t_3-\alpha_3 x_3)\right]} \\ \times e^{i\left[\beta_1 x_1 \epsilon_1^2 + \beta_2 x_2 \epsilon_2^2 + \beta_3 x_3 (-\epsilon_1-\epsilon_2)^2\right]}. \quad (9)$$

We introduce two new variables $t$ and $\tau$ defined in such a way that

$$t_2 = t_1 + t, \\ t_3 = t_1 + t + \tau = t_2 + \tau. \quad (10)$$

Thus $t$ is the delay between the detection of photon 2 and photon 1, while $\tau$ is the delay between photon 3 and 2. All of the Gaussian integrals can be then performed by substituting Eqs. (4) and (10) into Eq. (9) and using the identity

$$\int_{-\infty}^{\infty} dx e^{-(ax^2+bx+c)} = \sqrt{\frac{\pi}{a}} e^{(b^2-4ac)/4a}. \quad (11)$$

Evaluating the integrals gives a coincidence probability density $P(t,\tau) = \psi^*\psi$ that can be written in the form

$$P(t,\tau) = c'' Exp\left(2\sigma_F^2 \frac{N_1' + N_2' + N_3'}{D}\right). \quad (12)$$

Here

$$N_1' = -t^2\left[3 + 4\sigma_F^4\left((\beta_2 x_2)^2 + (\beta_2 x_2)(\beta_3 x_3) + (\beta_3 x_3)^2\right)\right], \\ N_2' = -t\tau\left[3 + 4\sigma_F^4\{(\beta_1 x_1)(\beta_2 x_2) + 2(\beta_2 x_2)^2 \\ + (\beta_2 x_2)(\beta_3 x_3) - (\beta_1 x_1)(\beta_3 x_3)\}\right], \\ N_3' = -\tau^2\left[3 + 4\sigma_F^4\left((\beta_1 x_1)^2 + (\beta_1 x_1)(\beta_2 x_2) + (\beta_2 x_2)^2\right)\right], \quad (13)$$

and

$$D = 9 + 8\sigma_F^4\left[2(\beta_1 x_1)^2 + 2(\beta_2 x_2)^2 + 2(\beta_3 x_3)^2 \\ + (\beta_1 x_1)(\beta_2 x_2) + (\beta_1 x_1)(\beta_3 x_3) + (\beta_2 x_2)(\beta_3 x_3)\right] \\ + 2\sigma_F^4\left[(\beta_1 x_1)(\beta_2 x_2) + (\beta_1 x_1)(\beta_3 x_3) + (\beta_2 x_2)(\beta_3 x_3)\right]^2\right]. \quad (14)$$

We have simplified the form of the equations by making the substitution $t_i \to t_i - \alpha_i x_i$, which subtracts off the effects of the group velocities. The calculations are discussed in more detail in the Appendix. These results will be plotted and discussed in Section III.

We have also considered the situation in which photon 3 is passed through a narrow-band filter before it is detected, as illustrated in Fig. 2. Post-selecting on a specific frequency $\omega_3 = \tilde{\omega}_3$ collapses the state of the system and effectively introduces a Dirac delta function $\delta(\omega_3 - \tilde{\omega}_3)$ into the integrals. Following a similar process as before, we arrive at a probability distribution for the detection times of photons 1 and 2 that is given by

$$P(t) = cExp\left[\frac{-(t-\bar{t})^2}{2\sigma_T'^2}\right], \quad (15)$$

where

$$\bar{t} = (\alpha_2 x_2 - \alpha_1 x_1) + (\beta_2 x_2 - \beta_1 x_1)\left(\frac{\omega_0}{3} - \omega_3\right) \quad (16)$$

and

$$\sigma_T'^2 = \frac{1/\sigma_F^4 + (\beta_1 x_1 + \beta_2 x_2)^2}{1/\sigma_F^2}. \quad (17)$$

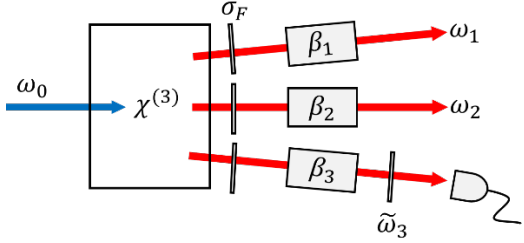

**Figure 2:** The same entangled three-photon source shown in Fig. 1, with the addition of a narrow-band filter and detector placed in the path of photon 3. This allows post-selection on the frequency of photon 3.

We note from Eq. (17) that the effects of dispersion on photons 1 and 2 can be eliminated if we choose $\beta_1 x_1 = -\beta_2 x_2$, as in Ref. [3]. In addition, it can be seen from Eq. (16) that the choice of the post-selected frequency $\tilde{\omega}_3$ can be used to control the relative detection times of the other two photons. All of these features are due to the entanglement of the third photon with the other two.

### III. Classical pulses

We now calculate the analogous results for the case of three classical pulses of light propagating in three separate media, such as three optical fibers. The correlated intensity distribution of the pulses after propagation will be compared to the results for three entangled photons as given in Eq. (12).

The electric field $E_i(0,t_i)$ of the classical pulses emitted at the source will be assumed to be Gaussians described by

$$E_i(0,t_i) = \frac{E_0}{2\pi} \int_{-\infty}^{\infty} e^{-(\omega_i - \omega_F)^2/2\sigma_F^2} e^{-i\omega_i t_i} d\omega_i. \quad (18)$$

After the pulses propagate through their respective media, the electric fields at the three detectors at $x_i$ become

$$E_i(x_i,t_i) =$$
$$\frac{E_0}{2\pi} \int_{-\infty}^{\infty} e^{-(\omega_i - \omega_F)^2/2\sigma_F^2} e^{i\left[k_F + \alpha_i(\omega_i - \omega_F) + \beta_i(\omega_i - \omega_F)^2\right]x_i} e^{-i\omega_i t_i} d\omega_i,$$
(19)

where $k_F = \omega_F / c$. Equation (19) can be integrated to give

$$E_i(x_i,t_i) = \frac{E_0}{2\pi^{1/2} a_i} \exp\left[-\frac{(t_i - \alpha_i x_i)^2 (\sigma_0^2 + i\beta_i x_i)}{4(\sigma_0^4 + (\beta_i x_i)^2)}\right], \quad (20)$$

where

$$a_i^2 = \frac{1}{2\sigma_F^2} - i\beta_i x_i, \quad (21)$$

and

$$\sigma_0^2 = \frac{1}{2\sigma_F^2}. \quad (22)$$

An irrelevant phase factor has been dropped.

Multiplying the fields in Eq. (20) by their complex conjugates give the intensities

$$I_i(x_i,t_i) = \frac{E_0^2}{4\pi |a_i|^2} \exp\left[\frac{-(t_i - \alpha_i x_i)^2}{2\sigma_i^2}\right] \quad (23)$$

where

$$\sigma_i^2 \equiv \frac{(\sigma_0^4 + \beta_i x_i)}{\sigma_0^2}. \quad (24)$$

If the intensities are sufficiently weak that single-photon detectors (or their classical equivalent) can be used, the detection probabilities at any given time will be proportional to the respective local field intensities. Thus the probability $P(t_1,t_2,t_3)$ of obtaining three detection events at times $t_i$ is

$$P(t_1,t_2,t_3) = \eta I_1(x_1,t_1) I_2(x_2,t_2) I_3(x_3,t_3), \quad (25)$$

where the constant $\eta$ is related to the detection efficiencies. The probability distribution $P(t,\tau)$ that pulses 2 and 3 are measured at time delays $t$ and $t+\tau$ after pulse 1 respectively is then given by integrating over $t_1$, which gives

$$P(t,\tau) = \eta \frac{E_0^6}{(4\pi)^3 |a_1|^2 |a_2|^2 |a_3|^2} \times \int e^{-\frac{1}{2}\left[\frac{(t_1-\alpha_1 x_1)^2}{\sigma_1^2} + \frac{(t_1+t-\alpha_2 x_2)^2}{\sigma_2^2} + \frac{(t_1+t+\tau-\alpha_3 x_3)^2}{\sigma_3^2}\right]} dt_1. \quad (26)$$

The effects of the group velocities can be ignored by making a change of variables as in Eq. (10). This simplifies Eq. (26), which can then be integrated to give

$$P(t,\tau) = cE \exp\left(2\sigma_F^2 \frac{N_1+N_2+N_3}{D}\right), \quad (27)$$

Here

$$\begin{aligned}
N_1 &= -t^2\left(1+2\sigma_F^4\left((\beta_2 x_2)^2+(\beta_3 x_3)^2\right)\right), \\
N_2 &= -t\tau\left(1+4(\beta_2 x_2)^2 \sigma_F^4\right), \\
N_3 &= -\tau^2\left(1+2\sigma_F^4\left((\beta_2 x_2)^2+(\beta_1 x_1)^2\right)\right), \\
D &= 3+8\sigma_F^4\left[(\beta_1 x_1)^2+(\beta_2 x_2)^2+(\beta_3 x_3)^2\right. \\
&\quad +2\sigma_F^4\left\{(\beta_1 x_1)^2(\beta_2 x_2)^2+(\beta_2 x_2)^2(\beta_3 x_3)^2\right. \\
&\quad \left.\left.+(\beta_3 x_3)^2(\beta_1 x_1)^2\right\}\right].
\end{aligned} \quad (28)$$

These results will also be plotted and compared to the entangled-photon case in the next section.

### IV. Comparison of the classical and quantum results

Fig. 3 compares the classical and quantum-mechanical timing distributions as calculated in sections II and III for an arbitrary choice of the relevant parameters where the filter bandwidth was relatively narrow ($\sigma_F = 0.10$). It can be seen that the effects of dispersion have not been completely cancelled out in the quantum mechanical results. Nevertheless, the timing uncertainties are significantly less than in the classical case due to dispersion cancellation.

Fig. 4. shows similar timing distributions as in Fig. 3 but with a relatively large filter bandwidth of $\sigma_F = 0.50$ and a different set of dispersion coefficients. The differences between the quantum-mechanical and classical cases are significantly larger than was the case for the smaller filter bandwidths in Fig. 3. As one might expect, the effects of dispersion and dispersion cancellation are larger for larger bandwidths.

The analytic calculations of section II assumed that the bandwidth $\sigma_F$ of the filters was much smaller than $\omega_0$. This condition is satisfied reasonably well in Fig. 3, where $\sigma_F = 0.10$ and $\omega_0 = 1.0$, but not as well in Fig. 4 where $\sigma_F = 0.50$. In order to assess the validity of this approximation, the analytic results based on the assumption that $\sigma_F \ll \omega_0$ were compared with the results of a numerical calculation where the range of integration was not extended to $-\infty$. The numerical results are shown in Fig. 5. It can be seen that the width of the probability distribution is somewhat underestimated in the analytic calculations, but that effect is much smaller than the difference between the classical and quantum-mechanical results in Fig. 4.

In the original case of two entangled photons [3], the quantum-mechanical dispersion was proportional to $(\beta_1 x_1 + \beta_2 x_2)^2$ while the classical dispersion was proportional to $(\beta_1 x_1)^2 + (\beta_2 x_2)^2$. This allowed the quantum-mechanical dispersion to be canceled nonlocally by choosing $\beta_1 x_1 = -\beta_2 x_2$, which has no effect on the classical dispersion. Complete dispersion cancellation would be possible for three entangled photons as well if the dispersion were simply proportional to $(\beta_1 x_1 + \beta_2 x_2 + \beta_3 x_3)^2$, but it can be seen from Eq. (13) that the quantum-mechanical dispersion also depends on terms such as $(\beta_1 x_1)^2$, which makes it impossible to cancel out all of the effects of dispersion nonlocally for three photons.

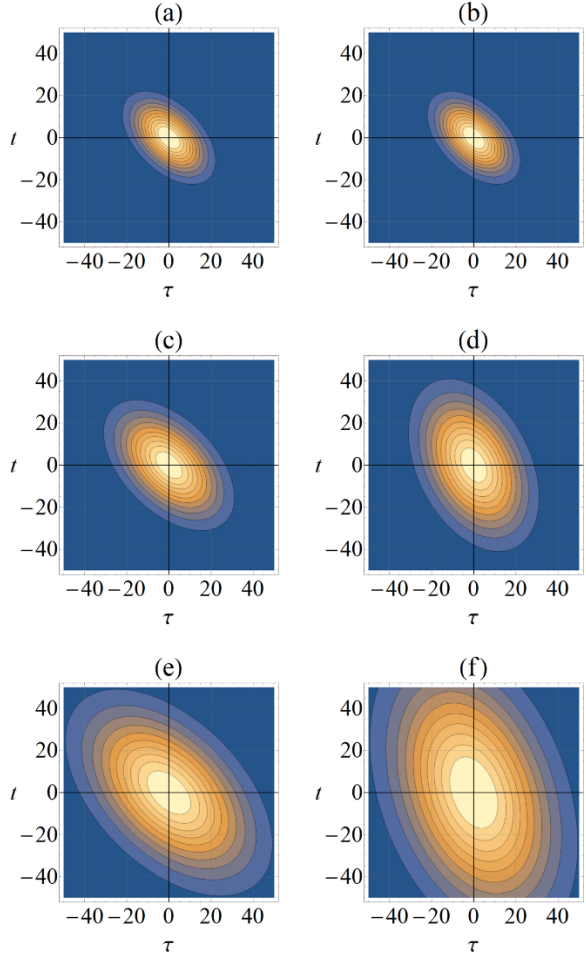
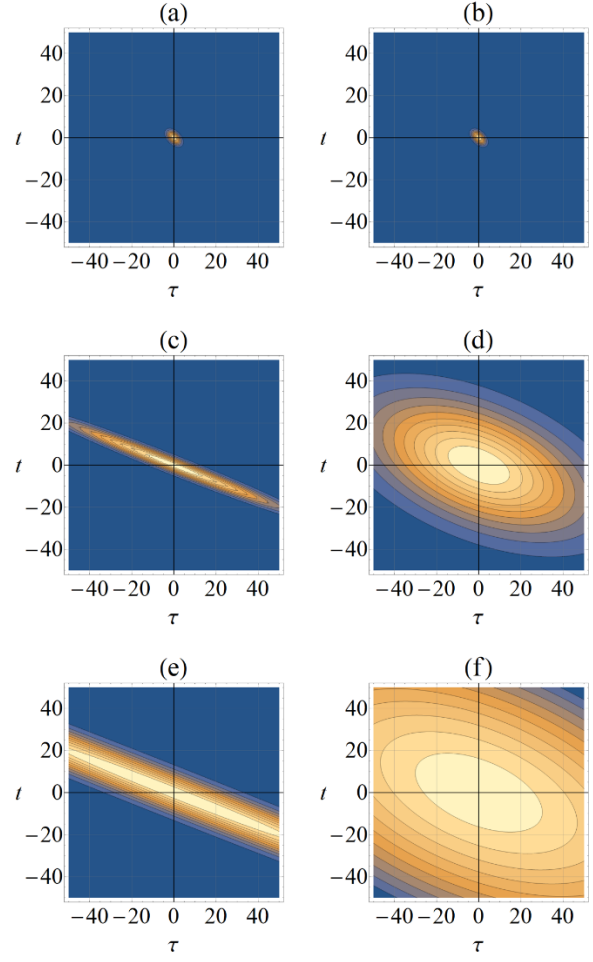

**Figure 3.** Probability distribution that detectors 2 and 3 will detect a single photon with a time lag of $t$ and $t + \tau$ (dimensionless) after detector 1, respectively, for a relatively narrow-band filter with $\sigma_F = 0.10$ (arbitrary units). Panels (a) and (b) correspond to the quantum mechanical and classical results, respectively, for no dispersion ($\beta_1 = \beta_2 = \beta_3 = 0$). The quantum and classical results are shown in (c, d) for a set of parameters with $\omega_0 = 1$, $\beta_1 x_1 = 100$, $\beta_2 x_2 = -50$, and $\beta_3 x_3 = -50$. Panels (e, f) correspond to $\beta_1 x_1 = 200$, $\beta_2 x_2 = -100$, and $\beta_3 x_3 = -100$. It can be seen that the timing uncertainties are significantly smaller for the quantum-mechanical results in (c, e) due to nonlocal cancellation of dispersion.

**Figure 4.** Probability distribution that detectors 2 and 3 will detect a single photon with a time lag of $t$ and $t + \tau$ (dimensionless) after detector 1, respectively, for a broader filter bandwidth ($\sigma_F = 0.50$) than in Fig. 3 (arbitrary units). The quantum-mechanical results are once again shown on the left-hand side while the corresponding classical results are on the right. Panels (a, b) correspond to no dispersion as before, while (c, d) correspond to $\omega_0 = 1$, $\beta_1 x_1 = 12.5$, $\beta_2 x_2 = -25$, and $\beta_3 x_3 = -37.5$. Panels (e, f) correspond to $\beta_1 x_1 = 50$, $\beta_2 x_2 = -100$, and $\beta_3 x_3 = -150$. The differences between the quantum-mechanical and classical results are more pronounced for a wider filter bandwidth.

Some of the terms can still be made to cancel in such a way that the quantum-mechanical dispersion is less than the corresponding classical dispersion, as can be seen in Figs. 3 and 4. This partial cancellation of dispersion may have useful applications in QKD, where the presence of an eavesdropper would increase the amount of dispersion. In addition, complete cancellation of dispersion can be obtained if we postselect on a specific value of the frequency of one of the photons as in Fig. 2, which may also have useful applications in quantum networks.

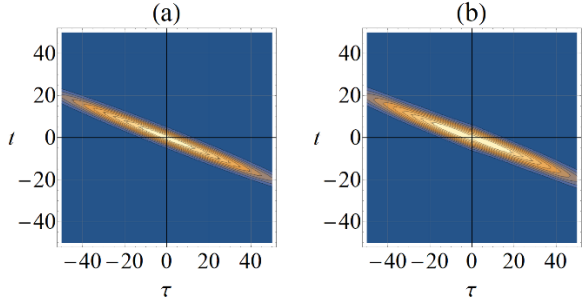

**Figure 5:** Comparison of the analytic and numerical calculations of the quantum-mechanical timing probability distribution as a function of the time delays $t$ and $\tau$ (dimensionless). (a) Analytic calculations using the narrowband-filter approximation. (b) Exact results from a numerical calculation. Both of these results correspond to the same parameters as in Fig. 4 (c). It can be seen that the width is somewhat larger in the numerical results but it is still much smaller than the corresponding classical results in Fig. 4.

## V. Extension to larger photon numbers

The previous results for three photons in a $\chi^{(3)}$ medium can be extended to $N$ photons in a $\chi^{(N)}$ medium in a straightforward way, since the calculations are based on a sequence of Gaussian integrals. To generalize from three-photons to $N$, we define the frequencies in terms of a set of parameters $\epsilon_i$ defined in such a way that

$$\omega_i = \frac{\omega_0}{N} + \epsilon_i. \qquad (29)$$

The energy conservation condition then becomes

$$\epsilon_N = -\sum_{i=1}^{N-1} \epsilon_i. \qquad (30)$$

Choosing filter functions with identical widths and making the approximation of narrow-band filters as before, we get an effective wave function of the form

$$\psi = c \int_{-\infty}^{\infty} \left( \prod_{p=1}^{N-1} d\epsilon_p \right) \left( \prod_{q=1}^{N} f_q \right) e^{i \left[ \sum_{r=1}^{N} (k_r x_r - \omega_r t_r) \right]}. \qquad (31)$$

This can be simplified to give

$$\psi = c' \int_{-\infty}^{\infty} \left( \prod_{p=1}^{N-1} d\epsilon_p \right) e^{i \left\{ \epsilon_q^2 \left( \beta_q x_q + \beta_N x_N + \frac{i}{\sigma_F^2} \right) - \epsilon_q \left[ (t_q - \alpha_q x_q) - (t_N - \alpha_N x_N) \right] \right\}}$$

$$\times e^{i \left[ \left( 2 \sum_{m<n}^{N-1} \epsilon_m \epsilon_n \right) \left( \beta_N x_N + \frac{i}{2\sigma_F^2} \right) \right]}. \qquad (32)$$

The integrals can all be performed but the results are lengthy. Nevertheless, it can be seen from Eq. (32) that the presence of the cross terms as before will prevent perfect cancellation of dispersion for $N > 2$. The dispersion cannot be completely cancelled for more than two entangled photons unless we postselect on the frequencies of all but two of the photons.

## VI. Summary and Conclusions

In summary, we have investigated the effects of nonlocal dispersion cancellation for three or more entangled photons. The analysis was based on a tripartite energy-time entangled state created directly by a single down conversion process in a nonlinear $\chi^{(3)}$ crystal. It can be shown that the same results can be obtained using two cascaded $\chi^{(2)}$ crystals if we ignore any dispersion between the first and second down-conversion crystals. The equations are identical in that case for narrow-band filters.

Our results show that nonlocal dispersion cancellation can reduce the width of the probability distribution for the coincidence events from a three-photon state as compared to the corresponding classical pulses. However, in general, complete dispersion cancellation cannot occur as it does for the two-photon case. This is a result of the fact that the dispersion is not simply proportional to $(\beta_1 x_1 + \beta_2 x_2 + \beta_3 x_3)^2$. The presence of other terms such as $(\beta_1 x_1)^2$ makes it impossible to completely cancel out all of the effects of dispersion nonlocally.

We also showed that postselecting on the frequency of one of the three photons does allow complete nonlocal dispersion cancellation for the remaining pair of photons. This effect is similar to the original two-photon case [3], except that the choice of the frequency in the postselection process can effectively control the difference in arrival times of the remaining pair of photons.

These effects may have practical applications in quantum communication protocols. The reduction in the timing uncertainties would allow the use of a smaller spacing between time bin qubits, with a corresponding increase in the data transmission rate. Quantum key distribution based on nonlocal dispersion cancellation between pairs of photons has already been proposed [24-27], and it may be possible to extend these techniques to larger numbers of photons in a network configuration. Dispersion cancellation has also been proposed as a means of increasing the imaging quality in biomedical applications [31-34] and for quantum clock synchronization [28-30].

Nonlocal dispersion cancellation for three or more photons is of fundamental scientific interest in addition to its potential applications, and these results will allow for future experimental investigations.

**Acknowledgements**

We would like to acknowledge valuable discussions with Richard Brewster and M. Brodsky. This work was supported in part by the Army Research Laboratory and by the National Science Foundation under Grant No. PHY-1802472.

**Appendix**

The Gaussian integrals in Eq. (9) can be evaluated by repeated use of Eq. (11). In order to simplify the results in the text, the effects of the group velocities were removed by making the substitution $t_i \to t_i - \alpha_i x_i$. The more general results including the group velocity are given in this Appendix.

In that case the coincidence probability is given by

$$P(t,\tau) = c'' Exp\left(2\sigma_F^2 \frac{N_1 + N_2 + N_3 + N_4 + N_5 + N_6}{D}\right).$$
(A1)

The value of $D$ is the same as in the text, while

$$N_1 = t^2\left[-3 - 4\sigma_F^4\left(B_2^2 + B_2 B_3 + B_3^2\right)\right],$$

$$N_2 = t\tau\left[-3 - 4\sigma_F^4\left(B_1 B_2 + 2B_2^2 - B_1 B_3 + B_2 B_3\right)\right],$$

$$N_3 = \tau^2\left[-3 - 4\sigma_F^4\left(B_1^2 + B_1 B_2 + B_2^2\right)\right],$$

$$N_4 = t$$
$$\left[A_3\left(3 + 4\sigma_F^4\left(B_2(B_1 + 2B_2) + (-B_1 + B_2)B_3\right)\right)\right.$$
$$-2A_1\left(3 + 4\sigma_F^4\left(B_2^2 + B_2 B_3 + B_3^2\right)\right)$$
$$\left.+A_2\left(3 + 4\sigma_F^4\left(-B_1 B_2 + (B_1 + B_2)B_3 + 2B_3^2\right)\right)\right],$$

$$N_5 = \tau$$
$$\left[2A_3\left(3 + 4\sigma_F^4\left(B_1^2 + B_1 B_2 + B_2^2\right)\right)\right.$$
$$-A_2\left(3 + 4\sigma_F^4\left(2B_1^2 - B_2 B_3 + B_1(B_2 + B_3)\right)\right)$$
$$\left.-A_1\left(3 + 4\sigma_F^4\left(B_1(B_2 - B_3) + B_2(2B_2 + B_3)\right)\right)\right],$$

$$N_6 =$$
$$-A_3^2\left(3 + 4\sigma_F^4\left(B_1^2 + B_1 B_2 + B_2^2\right)\right)$$
$$+A_2 A_3\left(3 + 4\sigma_F^4\left(B_1(2B_1 + B_2) + (B_1 - B_2)B_3\right)\right)$$
$$-A_2^2\left(3 + 4\sigma_F^4\left(B_1^2 + B_1 B_3 + B_3^2\right)\right)$$
$$-A_1^2\left(3 + 4\sigma_F^4\left(B_2^2 + B_2 B_3 + B_3^2\right)\right)$$
$$+A_1\left[A_3\left(3 + 4\sigma_F^4\left(B_2(B_1 + 2B_2) + (-B_1 + B_2)B_3\right)\right)\right.$$
$$\left.+A_2\left(3 + 4\sigma_F^4\left(-B_1 B_2 + (B_1 + B_2)B_3 + 2B_3^2\right)\right)\right].$$
(A2)

In order to shorten these expressions, we have used the notation that $B_i \equiv \beta_i x_i$ and $A_i \equiv \alpha_i x_i$. These results include the effects of the group velocities.